\documentclass[%
 preprint,
superscriptaddress,
nofootinbib,
 amsmath,amssymb,aps
]{revtex4-2}

\usepackage{amsmath}
\usepackage[section]{placeins}
\usepackage[utf8]{inputenc}
\usepackage[english]{babel}
\makeatletter\AtBeginDocument{\let\@elt\relax}\makeatother
\usepackage{float}
\usepackage{isotope}

\usepackage[dvipsnames]{xcolor}

\usepackage{soul}
\usepackage{graphicx}
\usepackage{dcolumn}
\usepackage{bm}

\usepackage{graphicx}
\usepackage{dcolumn}
\usepackage{bm}

\usepackage[draft, inline, nomargin]{fixme}
\fxsetup{theme=color, mode=multiuser}

\FXRegisterAuthor{bc}{1}{\color{red}BC}
\FXRegisterAuthor{ag}{2}{\color{blue}AG}
\FXRegisterAuthor{ph}{3}{\color{magenta}PH}

\newcommand\snowmass{\begin{center}\rule[-0.2in]{\hsize}{0.01in}\\\rule{\hsize}{0.01in}\\
\vskip 0.1in Submitted to the  Proceedings of the US Community Study\\ 
on the Future of Particle Physics (Snowmass 2021)\\ 
\rule{\hsize}{0.01in}\\\rule[+0.2in]{\hsize}{0.01in} \end{center}}

\newcommand{\CNP}{\affiliation{Center for Neutrino Physics, Physics Department, Virginia Tech, Blacksburg, VA}}

\newcommand{\VT}{\affiliation{Physics Department, Virginia Tech, Blacksburg, VA}}

\newcommand{\GT}{\affiliation{George W. Woodruff School of Mechanical Engineering,
Georgia Institute of Technology, Atlanta, GA, USA}}

\newcommand{\UNM}{\affiliation{University of New Mexico, Albuquerque, NM}
}

\newcommand{\UM}{\affiliation{Department of Nuclear Engineering and Radiological Sciences, University of Michigan, Ann Arbor, MI}}

\newcommand{\UZH}{\affiliation{University of Zurich, Switzerland}}

\newcommand{\LLNL}{\affiliation{Lawrence Livermore National Laboratory, Livermore, CA}}

\begin{document}

\title{Passive low energy nuclear recoil detection with color centers -- PALEOCCENE}

\author{Krystal Alfonso}\CNP
\author{Gabriela R. Araujo}\UZH
\author{Laura Baudis}\UZH
\author{Nathaniel Bowden}\LLNL
\author{Bernadette K. Cogswell}\CNP
\author{Anna Erickson}\GT
\author{Michelle Galloway}\UZH
\author{Adam A. Hecht}\UNM
\author{Rathsara R. H. Herath Mudiyanselage}\VT
\author{Patrick Huber}\email{pahuber@vt.edu}\CNP
\author{Igor Jovanovic}\UM
\author{Giti A. Khodaparast}\VT
\author{Brenden A. Magill}\VT
\author{Thomas O'Donnell}\CNP
\author{Nicholas W. G. Smith}\VT
\author{Xianyi Zhang}\LLNL

\date{\today}%

\maketitle
\snowmass{}

\newpage
\section{Executive Summary}

The PALEOCCENE concept~\cite{Cogswell:2021qlq} offers the potential for room-temperature, passive and robust detectors in the gram to kilogram range for the detection of low-energy nuclear recoil events. Nuclear recoil events can be caused by neutron scattering, coherent elastic neutrino nucleus scattering (CEvNS) or dark matter scattering and therefore, PALEOCCENE could find applications in all three areas. Nuclear recoils result in damage to the crystal lattice and some of these damage sites can become optically active, so-called color centers. The formation energy for these damage sites is in the range of a few tens of electronvolt and sets the intrinsic nuclear recoil detection threshold. In order to be sensitive to rare low-energy events optical detection of the fluorescence of single color centers in bulk volumes is envisaged by using so-called light sheet microscopy.
Color centers are known to occur in a wide range of materials and thus, it appears plausible that a range of materials could be suitable for the PALEOECCENE approach. A multi-disciplinary collaboration of experts in particle, solid state and nuclear physics, as well as dark matter detection and nuclear engineering has started an  R\&D program  to investigate the feasibility of this concept. By using neutrons as projectiles to induce nuclear recoils around 100\,keV our measurements indicate an agreement with previous results on CaF$_2$~\cite{Mosbacher:2019igk} on radiation induced photo luminescence at much lower dose rates. We also observed that this signature remains stable over a period of weeks in crystals kept at room temperature. Future plans include the demonstration of correlating bulk photo luminescence with microscopic imaging of color centers and radiation dose and to establish the level of (in-)sensitivity to gamma and beta radiation. Eventually, these studies will need to be extended to lower recoil energies in the range relevant for applications in dark matter and reactor CEvNS experiments. The studies with neutrons can be complemented by using low-energy ions (keV) which allow for more controlled experiments. More detailed studies of the signal persistence under a range of conditions and in various materials are foreseen. Studies to understand the dependence of the sensitivity on the crystal direction are motivated by  simulation results~\cite{MORRIS2020109293}, which indicate that there should be a significant anisotropy in crystal response. If experimentally confirmed this would allow for directional detectors.

We discuss several R\&D needs for demonstration of the PALEOCCENE concept, which would benefit neutron, dark matter and CEvNS detection.

\newpage
\section{Introduction}
\label{sec:intro}

Observing known and finding new phenomena at ever lower energies is one of the frontiers in particle physics today. Nuclear recoil is one such signature at low energy and can arise from the interaction of a wide range of particles, ranging from neutrons, ions, over neutrinos to dark matter~\cite{drukier,Goodman:1984dc,Battaglieri:2017aum}. There are many well established detection technologies for nuclear recoils usually exploiting ionization, scintillation or phonons or a combination thereof.  An alternative of using crystal damage  was proposed for dark matter searches~\cite{Essig:2016crl} and subsequently explored both for damage tracks~\cite{Drukier_2019,Edwards_2019} and point defects~\cite{Budnik:2017sbu,Rajendran2017}. This appears particularly interesting, given that conventional techniques seem to experience unexplained backgrounds at recoil energies of less than 1,000\,eV~\cite{Proceedings:2022hmu}, presenting a major obstacle for both searches for sub-GeV dark matter and the observation of coherent elastic neutrino nucleus scattering (CEvNS) at reactors.

\begin{figure}
    \centering
    
     \includegraphics[width=0.5\textwidth]{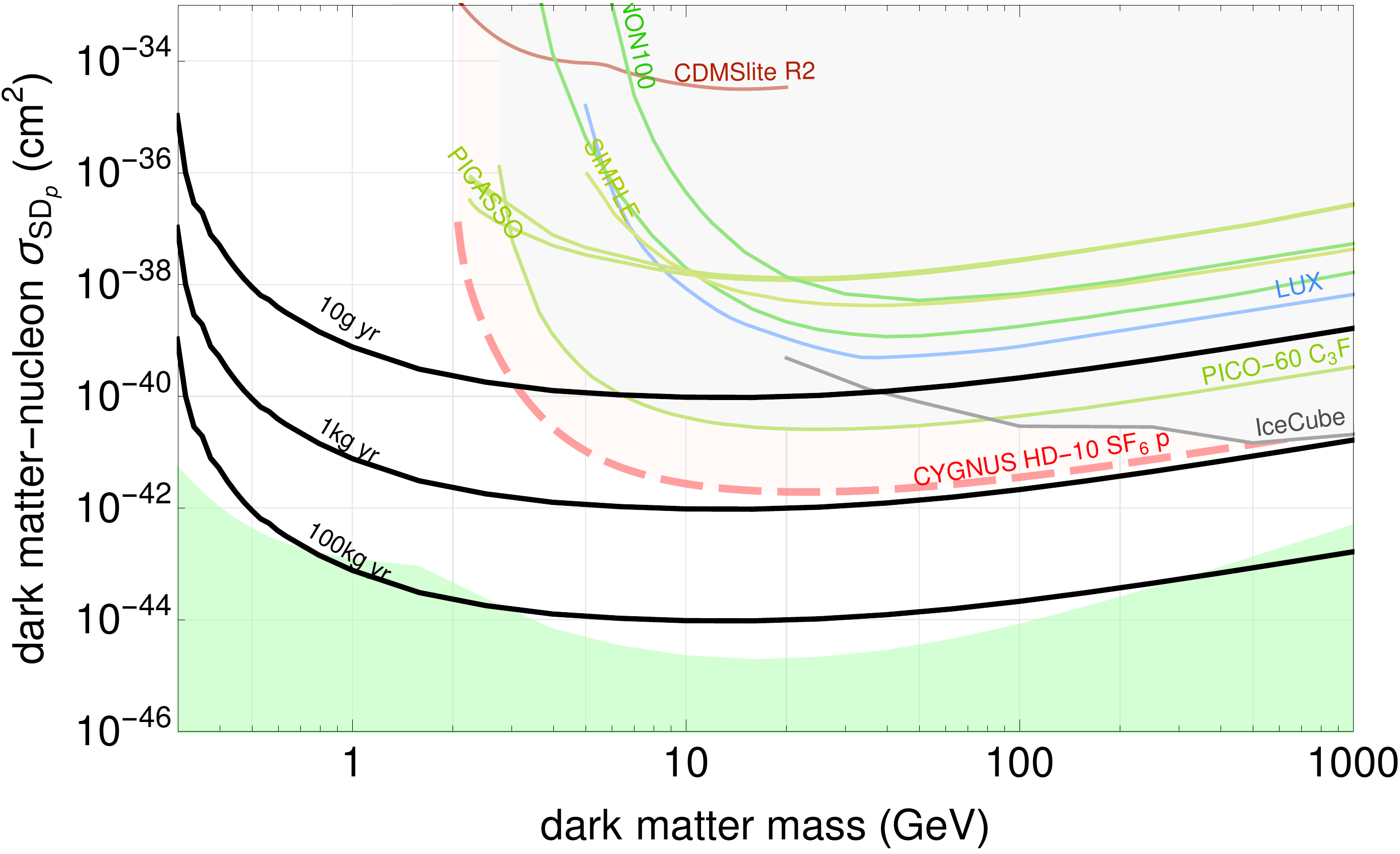}%
     \includegraphics[width=0.5\textwidth]{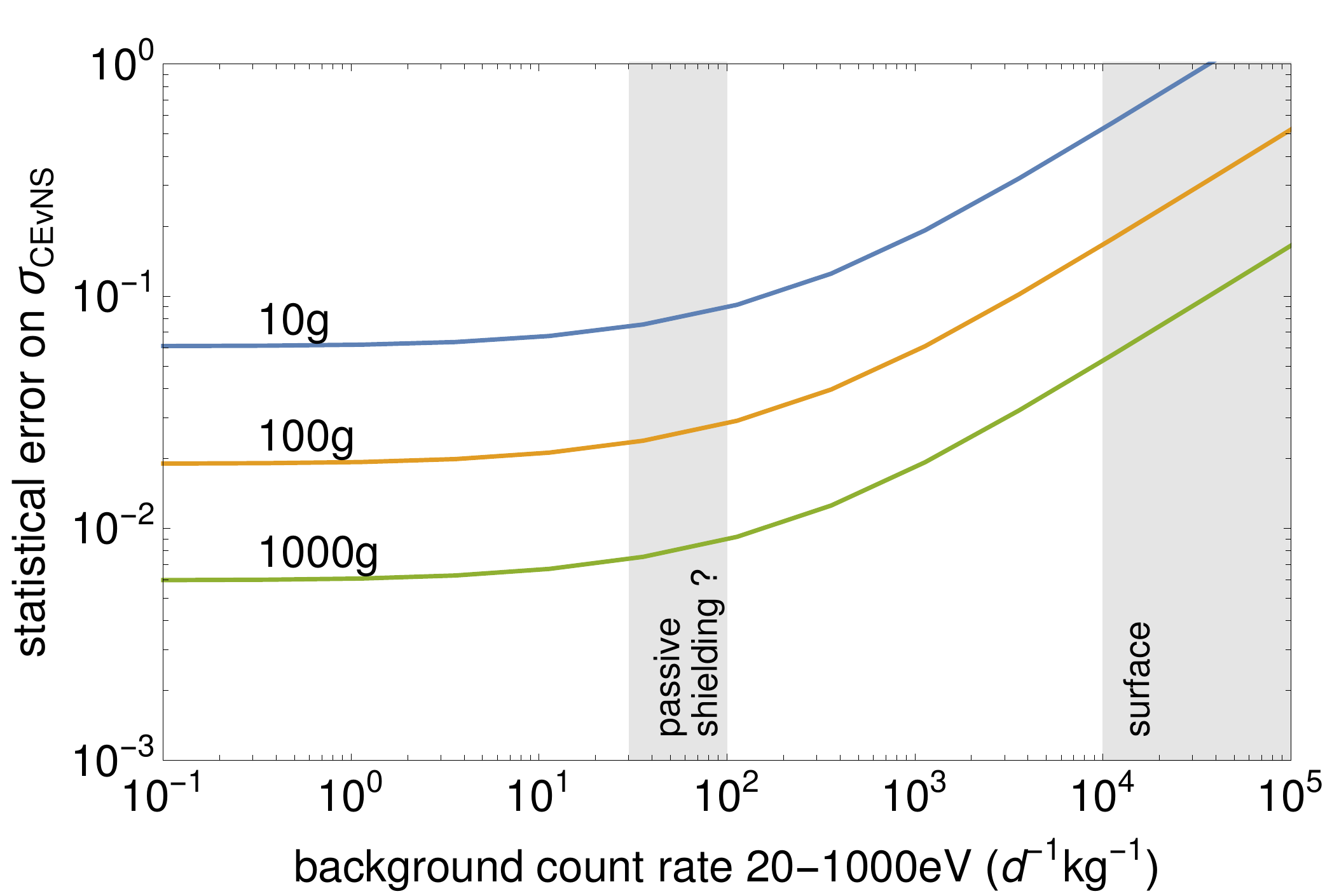}
    \caption{Left hand panel: Shown is the 95\% CL sensitivity with LiF for spin-dependent dark matter scattering (black lines) for different exposures assuming a background free experiment. The green shaded region is the neutrino floor for LiF. All other experiments and exclusion limits have been plotted using~\cite{dmplot}, specifically SIMPLE II~\cite{Felizardo2012}, PICO-60~\cite{Behnke2017}, LUX~\cite{Akerib_2017}, Panda-X II~\cite{Fu2017}, XENON-100~\cite{Aprile2016}, CDMSlite~\cite{cdmslite}. Right hand panel: Shown is the 1\,$\sigma$ error on the CEvNS cross section obtainable at a nuclear reactor for a variety of detector masses in a vacancy-based counting analysis, as a function of the neutron background rate. The exposure is one year at 20\,m from a 3\,GW$_\mathrm{th}$ reactor. Figures and captions taken from Ref.~\cite{Cogswell:2021qlq}.}    
    \label{fig:sensitivities}
\end{figure}

In this note we will outline the R\&D required to eventually have working detectors based on a specific class of radiation induced crystal defects, so called color centers. We term this approach PAssive Low Energy Optical Color CEnter Nuclear rEcoil (PALEOCCENE) detection. Color centers have been known for more than 100 years and more recently are studied in quantum information science~\cite{Doherty_2013} and metrology~\cite{nv}, since they can be individually detected by fluorescence spectroscopy. In Ref.~\cite{Cogswell:2021qlq} an optical, non-destructive readout scheme based on so-called light sheet microscopy, also known as selective plane illumination microscopy (SPIM)~\cite{lightsheet} was proposed. SPIM allows in principle to scale this technique to kilogram quantities of detector material. 
The resulting detectors would be passive and operate at room-temperature. They could find application in direct dark matter searches, CEvNS detection of reactor neutrinos for both basic science and nuclear security as well as for the detection of neutrons. The resulting sensitivities are shown in Fig.~\ref{fig:sensitivities}.

\section{Detection concept}
\label{sec:concept}

A recoiling nucleus will lose all its kinetic energy to the host crystal lattice via phonons, ionization, and lattice defect creation. Where the energy transfer to a crystal atom exceeds the binding energy, the atom can be dislodged and move through the crystal. This site can remain unoccupied, leading to a vacancy, or it can be filled by the initial projectile or some other dislocated atom.  A permanent vacancy forms when the atom is removed far enough from its lattice site so that recombination is no longer possible. This distance is on the order of 1--2\,nm. Since the stopping power for most ions is around 20--100\,eV/nm, an approximate energy for the recoiling nucleus in the range 20--200\,eV is required for the creation of a permanent vacancy. This energy is, at best, indirectly related to the binding energy of a given solid or the defect formation energy. Therefore, it makes sense to introduce the concept of a threshold damage energy (TDE), with the idea that below the TDE no permanent vacancies are formed. 

\begin{figure}
   \centering
    \includegraphics[width=0.5\textwidth]{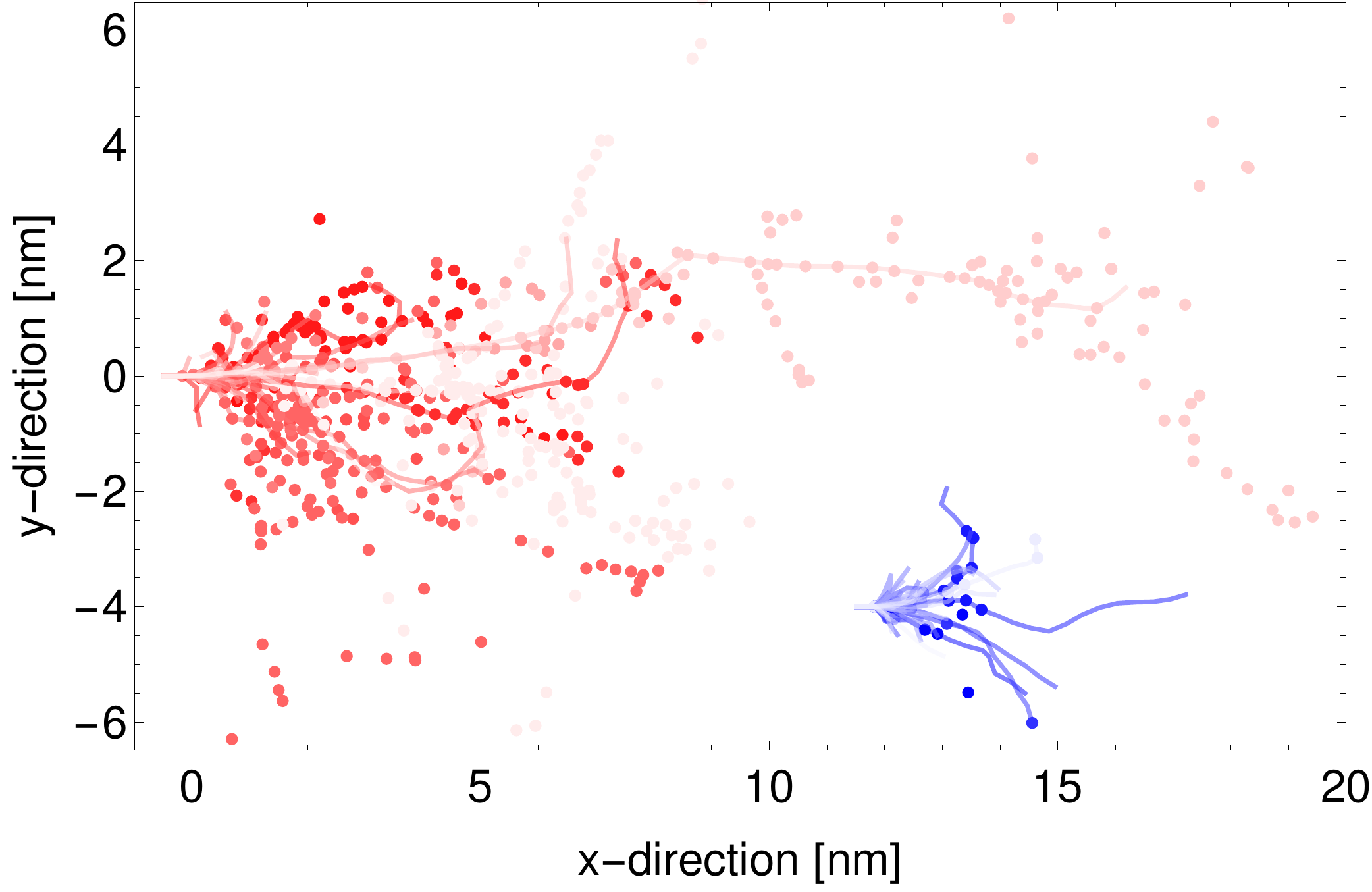}
    \caption{Shown is the overlay of 50 typical cosmic ray neutron (red) and reactor CEvNS events (blue) in NaI. Vacancies are marked by disks and tracks created by the primary recoil are marked by a line. Figure and caption taken from Ref.~\cite{Cogswell:2021qlq}.}
    \label{fig:events}
\end{figure}

In Fig.~\ref{fig:events} we show a set of 50 cosmic ray neutrons (red) and reactor CEvNS events (blue) in NaI, each. The lines are the tracks caused by the primary recoil and the dots are the vacancies created. Neutron events, on average, are much larger spatially than CEvNS events since the mean recoil energy is large due to the $1/E$ neutron spectrum, whereas CEvNS events on average are only a few nm across since the reactor antineutrino spectrum ends at around 8\,MeV. The rate of vacancy formation depends on the TDE. The coherent neutrino cross-section is proportional to the atomic mass squared, $A^2$, whereas, neutron scattering cross sections are essentially independent of $A$. For CEvNS, this leads to a preference for materials with a large mass fraction of high-$A$ elements.

In terms of event rates, the sea-level neutron background from cosmic rays is about one order of magnitude larger than the typical CEvNS signal close to a power reactor in a high-A target like iodine or cesium. Also, the magnitude of neutron background changes relatively little as a function of the number of vacancies created, whereas, the magnitude of the CEvNS signal falls steeply with either the number of vacancies or track length. The flatness of the fast neutron background is a consequence of scattering kinematics combined with the approxmately $1/E$ neutron energy spectrum. This indicates that moderate shielding combined with a focus on events with few vacancies can help to control neutron backgrounds on a statistical basis. To obtain intrinsic rejection of gamma and beta radiation backgrounds, we implicitly assume that gammas and electrons do not cause  vacancy formation. Materials have been identified where optically active vacancies result predominantly from nuclear recoil events~\cite{Mosbacher:2019igk}. 

Color centers, in particular F-centers, have been proposed as a detection mechanism for direct dark matter searches~\cite{Budnik:2017sbu}. An F-center is a vacancy of the anion in an ionic crystal that traps an electron. The energy levels of this electron vacancy can be excited by light of a suitable wavelength and the subsequent de-excitation results in the emission of photons. Individual color centers can be detected using confocal microscopy. The limitation of this method is that when scanning a 3D volume only one voxel can be acquired at a time, so the total volume throughput is low.

\begin{figure*}
    \centering
    \includegraphics[width=\textwidth]{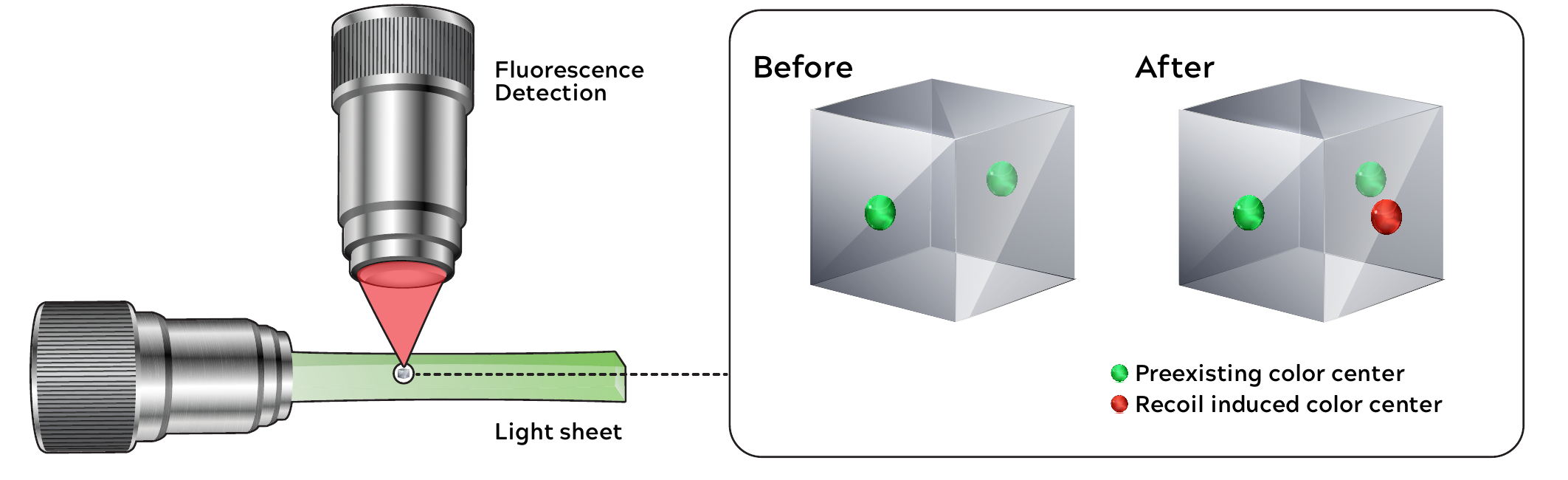}
    \caption{Schematic layout of light-sheet microscopy used for imaging single color centers pre- and post-exposure. Figure and caption taken from Ref.~\cite{Cogswell:2021qlq}.}
    \label{fig:schematic}
\end{figure*}

We propose the use of light-sheet microscopy, also known as Selective Plane Illumination Microscopy (SPIM)~\cite{Huisken1007}, an evolution of confocal microscopy. A sheet of light illuminates the entire field of view of the fluorescence light path. Thus, as many voxels as there are pixels in the camera along the fluorescence light arm can be imaged simultaneously, see Fig.~\ref{fig:schematic}. 

The required laser power for scanning depends sensitively on the density of pre-exisiting color centers and the details of the optical system (pixel size, resolution, numerical aperture of the objective etc.) but is estimated to be manageable as well as the associated heating of the sample, see Sec.~\ref{sec:tech} The crystal can be imaged prior to deployment and, thus, a voxel-by-voxel subtraction of background is feasible, as shown in Fig.~\ref{fig:schematic}. 

Existing studies and physics considerations help identify material candidates. Acceptable candidates have the following properties: high melting point, electrical insulator, permit optically active defects (color centers) pre-dominantly formed by nuclear recoils, available in optical quality crystals, contain high-mass elements for CEvNS, and have a low TDE. Color centers, i.e., crystal defects that are optically active, are common and are known to exist in almost all chemical group I-VIII compounds. However, color centers are not confined to this group of compounds and are found in many materials, including glasses. For LiF and BaF$_2$ color centers have been found confirmed to be created  much more easily by neutron than gamma irradiation~\cite{Mosbacher:2019igk}. That is, LiF and BaF$_2$ detectors will be relatively insensitive to ionizing radiation. Also, color centers in NaI, CsI, CaWO$_4$, BGO have been observed in response to X-ray and electron irradiation.

The neutron cross section is more or less independent of the atomic mass, $A$, whereas, the CEvNS cross section scales with $A^2$. Therefore, heavy elements are preferred. Within the group I-VIII compounds, this makes NaI and CsI prime candidates. Therefore, to obtain the largest possible number of events, we look for the heaviest possible elements and materials with the largest possible concentration of those heavy elements. In a practical detector this has to be balanced with the much lower recoil energy of heavy targets. Two candidates which optimized interaction rates are $\mathrm{Ca}\mathrm{W}\mathrm{O}_4$ and $\mathrm{Bi}_{12}\mathrm{Ge}\mathrm{O}_{20}$ (BGO). Both materials appear in nuclear and particle physics applications as scintillators and/or low-temperature bolometers. This implies their ready availability (and relative affordability) as large (100s of grams) crystals with good optical qualities and a low radioactive impurity content.

\section{R\&D stages}
\label{sec:rnd}

In order to have a detection technology which is relevant for applications in dark matter detection, reactor CEvNS, and neutron detection, we need to achieve:

\begin{enumerate}
    \item nuclear recoil threshold of $500$\,eV or less
    \item detector masses of $10$\,g or more
    \item accounting and rejection of ionizing backgrounds
    \item materials with high atomic mass and small atomic mass
\end{enumerate}

These requirements translate into specific physics goals. To reach a low threshold sensitivity, essentially single color center detection will be required given that the formation energy of color centers is of order 20-50\,eV. In order to be able to use detectors with a mass of 10\,g or more, scan speeds of order a 1\,cm$^2$ per day will be needed. Dealing with ionizing backgrounds ideally would be achieved by identifying the right material and wavelengths combination for color centers which have a small probability being formed by ionizing radiation, but a high probability for resulting from nuclear recoil, some candidate materials have been reported~\cite{Mosbacher:2019igk}. Having the ability to select the atomic mass and nuclear spin properties of detector materials is crucial for dark matter searches where for instance the choice of LiF would provide excellent sensitivity low-mass dark matter with spin-dependent interactions. On the other hand, for CEvNS experiments a high atomic mass greatly enhances the signal and co-deployment of high and low atomic mass detectors allows to account for neutron backgrounds. For each of these questions R\&D studies are needed, many of which provide information for more than one requirement. Also, expanding the range of suitable target materials is an important open question since the complexity of the optical system somewhat depends on the employed wavelengths, {\it e.g.} a system operating at visible wavelengths is generally significantly easier to realize than one operating in the deep UV. For different applications different nuclear properties are attractive and thus having a wide range of suitable materials would allow to tailor the material to the specific application.

As in all low-energy recoil experiments reaching the necessary signal sensitivity and threshold is only one part of the challenge with the other one being backgrounds. The PALEOCCENE approach is by design only sensitive to nuclear recoil backgrounds, which typically are caused by neutrons. There are direct cosmic ray neutrons and muon induced neutrons\footnote{Muon events inside the detector are in comparison much rarer and would result in large damage cascades with very different characteristics from the signal events.}. The total size of  the neutron background and ratio of direct versus muon induced neutrons is a steep function of the overburden. For the muon induced neutrons, the fact that PALEOCCENE is an off-line technology implies that a muon veto will not reduce this background. It appears however that the resulting background can be managed for close-to-the-surface deployments as for instance for reactor CEvNS searches. However, there is considerable uncertainty about the actual root cause for backgrounds below 1\,keV nuclear recoil energy: the vast majority of all experiments with sensitivity in this energy range experience a steep (exponential) increase in background beyond what a simple neutron recoil model would predict~\cite{Proceedings:2022hmu}. The fact that PALOECCENE is based on the observation of lattice defects which are a direct result of nuclear recoil  likely will make this technique respond differently to these unknown causes of background.

\subsection{Single color center sensitivity}

Sensitivity to single color centers can be achieved  by fluorescence microscopy~\cite{Haussler2020}, where a color center appears as an unresolved point source and can be identified by its excitation and emission wavelengths. So the first step has to be to establish  these excitation and emission wavelengths and to demonstrate that the corresponding color centers are indeed caused by nuclear recoils. Neutrons can provide a convenient source of nuclear recoils over a wide range of energies, with the  mean recoil energy $\bar E_r$ being given by
\begin{equation}
\bar E_r = \frac{2A}{(A+1)^2} E_n\,,
\end{equation}
with $E_n$ being the incident neutron energy and $A$ the atomic mass. For instance, in the case of calcium $A\simeq40$ and for 4.4\,MeV neutrons which is typical for an AmBe source, we find $E_r\simeq 200\,$keV. 

Using neutrons from a $^{252}$Cf source experiments to establish the relationship between neutron irradiation and fluorescence have been conducted~\cite{Mosbacher:2019igk} and identified a list of candidate materials which demonstrate insensitivity to ionizing radiation as well.  Calcium fluoride, CaF$_2$, is promising for follow up studies for the following reasons: its excitation and emission wavelength both are in the visible with 410\,nm and 580\,nm respectively; there is good theoretical understanding of its threshold damage energy, which has been determined by molecular dynamics simulations~\cite{MORRIS2020109293}; the atomic mass is intermediate; it is easily available in optical quality from vendors.

\begin{figure}[H]
    \centering
    \includegraphics[width=0.75\textwidth]{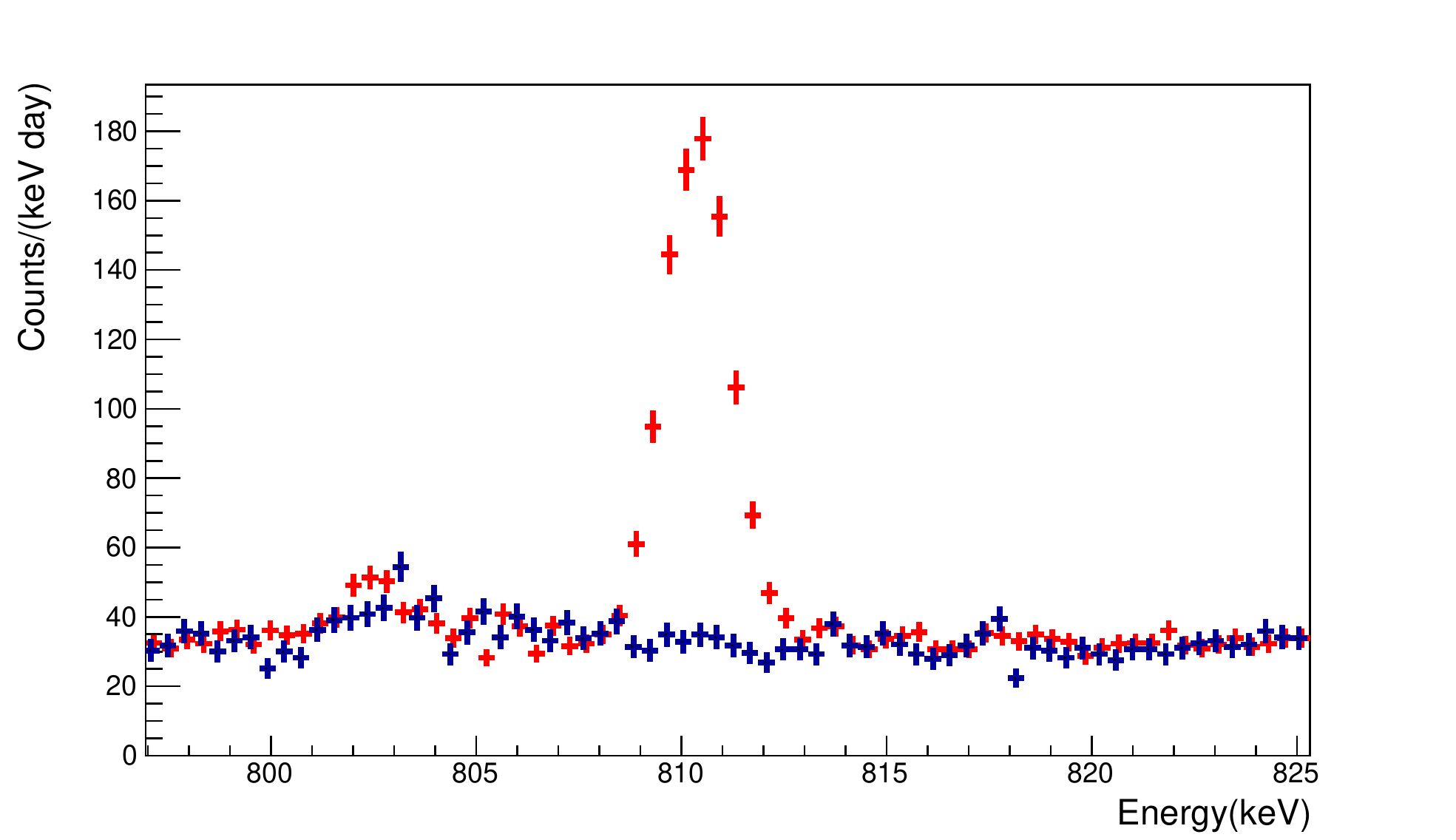}
    \caption{Spectrum observed with HPGe counter in the region of interest for $^{58}$Co $\gamma$ rays. (Blue) Ni neutron witness foil prior to neutron exposure; (Red) the same Ni foil after neutron exposure.}
  \label{Fig-HPGe-Ex}
\end{figure}

In order to establish a relationship between spectroscopic signals and neutron dose, CaF$_2$ samples have been exposed to neutrons from  a 10\,mCi AmBe source at a fluence of up to $10^8\,\mathrm{n}\,\mathrm{cm}^{-2}$,  for comparison in Ref.~\cite{Mosbacher:2019igk} a neutron fluence of $\sim 10^{10}\,  \mathrm{n}\,\mathrm{cm}^{-2}$ was used.  These relatively high neutron fluences ensure that a bulk spectroscopic response can be recorded and we could observe a clear spectroscopic response at a fluence of  $10^8\,\mathrm{n}\,\mathrm{cm}^{-2}$. Nickel foils were exposed with the samples to monitor the neutron dose via the yield of $^{58}$Co ($T_{1/2} \simeq 70$~d).  The $^{58}$Co yields were determined after the exposure by measuring the $\sim 810$ keV $\gamma$ ray emission rate from $^{58}$Co decay with a moderately shielded high purity germanium (HPGe) detector, see Fig.~\ref{Fig-HPGe-Ex}. For the foil geometries employed, the HPGe detector efficiency was estimated to be $(5.1 \pm 0.4)\%$. 
The samples were also imaged with a confocal microscope in order to identify a correlation between clusters of color centers and neutron dose. However, the volume which can be imaged in a reasonable time with a confocal microscope is limited, in our case approximately $2\times10^{-8}\,\mathrm{cm}^3$ and we expect about one neutron scattering event per $10^8\,\mathrm{n}\,\mathrm{cm}^{-2}$ in this volume. Therefore, a new set of samples was irradiated with neutrons from a significantly stronger AmBe source with fluences in the range $(2-30)\times 10^9\,\mathrm{n}\,\mathrm{cm}^{-2}$. The certified neutron output of the AmBe source is $1.3\times 10^8 \pm6\%\,\mathrm{n}\,\mathrm{sec}^{-1}$, with calibration conducted on January 5, 1966. The analysis of these data is ongoing and eventually should yield the observation of  color centers in confocal microscopy and establish a dose/signal size relationship.

\subsection {Far Field Optical Measurements}

In order to probe the emission spectrum in our CaF$_{2}$ samples, we detected the photoluminescence (PL) where the samples were excited by a continuous wave laser at 488 nm (Coherent Sapphire 488-20) with an output power of 25 mW at the sample. The PL was collected using a 0.55 m focal length spectrometer (HORIBA Jobin Yvon
iHR550) with a 0.1 mm slit width and a 150 grooves/mm, 500 nm
blaze grating. The PL was recorded by a liquid nitrogen cooled,
charge-coupled device (HORIBA Jobin Yvon Symphony II).

\begin{figure}[H]
    \centering
    \includegraphics[width=0.5\textwidth,trim={0cm 2.2cm 4cm 1.2cm}]{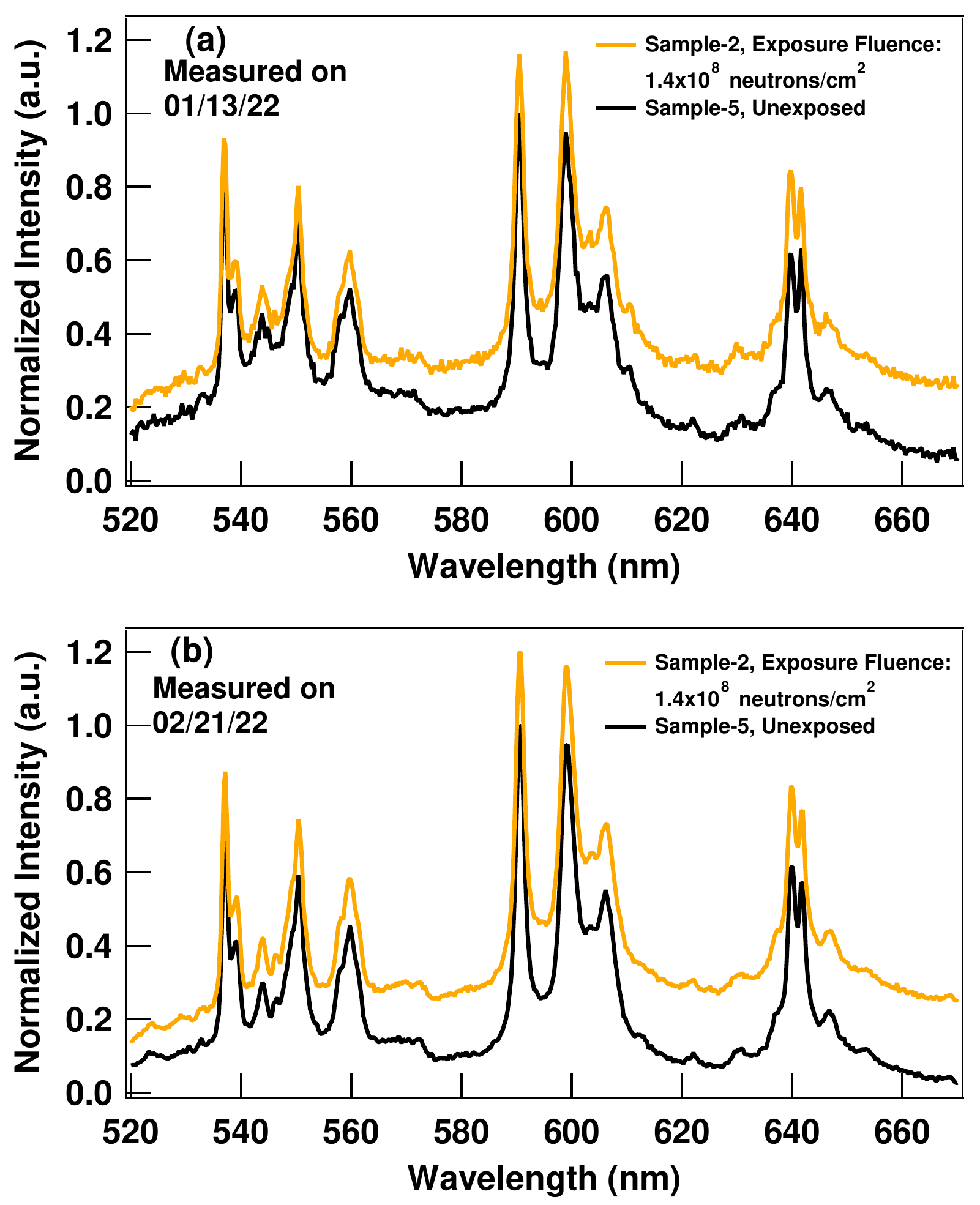}
    \newline
     \newline
     \newline
    \includegraphics[width=0.487\textwidth,trim={0cm 2.5cm 4cm 0cm}]{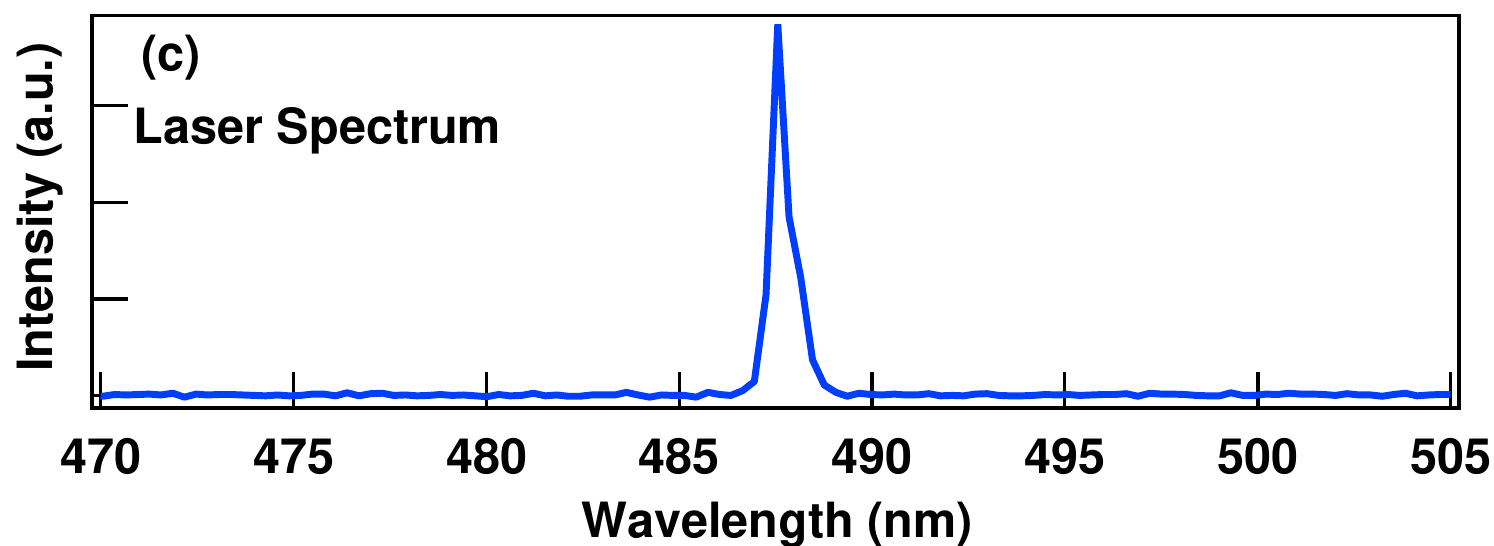}
    \newline
    \newline
    \caption{(a) Comparison of the PL emissions for two identical CaF$_{2}$ samples, in which one was irradiated (Sample-2) with a fluence of 1.4$\times$10$^8$ neutron/cm$^2$. The traces are the average of the PL from three different spots, for each sample. (b) The PL emissions under the same experimental conditions of panel(a); at a later time, to establish that the observations are reproducible. The traces here are the average of the PL at five different spots, for each sample. We should note, the PL's collection time in panel(a) was 5 seconds compared to 10 seconds; in panel(b), and this fact explains a better signal to noise ratio in the panel(b).
    (c) In this panel, we show the laser spectrum using the same grating; employed for the measurements on the CaF$_{2}$ samples. We kept the spectrometer slit almost closed for this measurement.}
  \label{Fig-PL}
\end{figure}

We selected two identical CaF$_{2}$ pieces described earlier, in which one piece (refered to as Sample-5) was kept as a reference sample, to be compared with another piece (refered to as Sample-2) which was irradiated at a fluence of 1.4$\times$10$^8$ neutron/cm$^2$. As shown in Fig. \ref{Fig-PL}(a-b), we are comparing the PL emissions of the two CaF$_{2}$ samples, where the observed changes, as a result of the neutron irradiations in Sample-2, have been reproducible after more than a month of the neutron exposure. In Fig. \ref{Fig-PL}(c) we present the laser emission profile, employed for the PL measurements. The laser beam size on the samples was about 0.5\,mm in diameter.

\subsection{Light sheet microscopy imaging}
\begin{figure}
    \centering
    \includegraphics[width=0.85\textwidth]{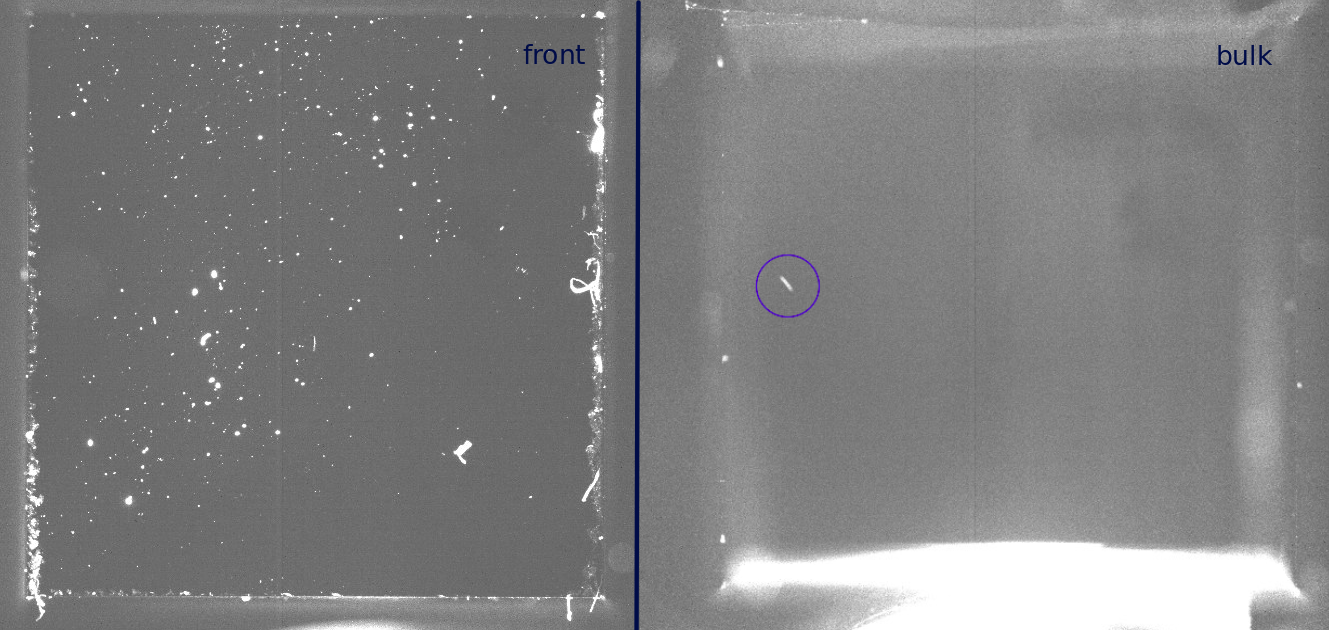}
    \caption{Images of surface (left) and bulk (right) scans of the irradiated CaF$_{2}$ cubic crystals.  }
    \label{fig:CaF2_AmBe_mic_image}
\end{figure}

We imaged two cubic CaF$_{2}$ crystals with a mesoSPIM light-sheet microscope ~\cite{mesoSPIM}, available at the Center for Microscopy and Image Analysis of the University of Zurich (UZH). This microscope is able to scan large (cm-sized) samples with a remarkable scan speed (~5\,min/cm$^3$) and near-isotropic resolution of a few micrometers in the x, y, z axes ~\cite{mesoSPIM}. The crystals were 1\,cm$^3$ large, transparent to visible light and had all sides polished. One of them was irradiated for 37.6 days with an AmBe source, with an activity of 0.5\,mCi, and yield of approximately 1100 n/s. To image the entire volume of the crystal, 1100 x-y scans without magnification and a 10\,$\mu m$ z-step size were taken. The excitation light had a wavelength of 405\,nm, provided by a 100\,mW laser coupled to a system that provides the light sheet ~\cite{mesoSPIM}. An emission filter that blocks the excitation light is used such that only fluorescence light is detected. A scan of the surface, as well as one of the bulk of the irradiated crystal are show in Fig. ~\ref{fig:CaF2_AmBe_mic_image}. 

In Fig. ~\ref{fig:CaF2_AmBe_mic_image}, we see many fluorescence spots on the surface of the crystal, likely dust or color centers produced by the mechanical stress of cutting and polishing the crystal. The bulk was mostly featureless at a first analysis, apart from one "track-like" spot observed at 7\,mm z-depth inside the crystal. Such a feature was not observed in the non-irradiated crystal. A detailed analysis of the data is still ongoing and further irradiation of crystals and subsequent scans with the mesoSPIM are planned\footnote{The UZH collaborators are registered users of the microscope center, with training and access to the mesoSPIM and other fluorescence microscopes.}. The aim of the next measurements is to firstly check the suitability of the open-source light-sheet mesoSPIM microscope for single color center imaging and secondly provide a relation between irradiation dose and color center  production.

\subsection{Persistence of color centers}

A critical question is the persistence of the color centers. Previous experiments on CaF$_2$ kept irradiated samples cool and in the dark to avoid both bleaching and annealing~\cite{Pelka:2017}. Bleaching occurs when too much light causes the color center to loose its optical activity, 
this also can be caused by ionizing radiation. Annealing is a process whereby crystal defects heal when the  temperature is sufficiently high. We will need to quantify both the light and temperature sensitivity of the optical activity of the color centers and explore materials which show persistent color centers at room temperature.

\subsection{Low-energy recoil studies}

To achieve nuclear recoils below 1\,keV, neutrons of around 20\,keV are needed. Suitable low-energy neutron sources can be realized by exploiting low-lying anti-resonances in materials like iron~\cite{Barbeau:2007qh} or scandium; another choice could be photo-neutron sources~\cite{Collar:2013xva} or the use of the reaction Li$^7$(p,n)Be$^7$ which results in $\sim 20\,$keV neutrons~\cite{Martin-Hernandez:2016ysa}. These types of neutron sources are also used in the study of quenching for dark matter experiments to characterize the detector response, see for instance Ref.~\cite{SuperCDMS:2022nlc}. Similar studies will be needed to eventually characterize the low-energy response of color center based detectors. To push to even lower recoil energies it was recently proposed to use the nuclear recoil following capture of thermal neutrons~\cite{Thulliez_2021,Villano:2021eof} which can yield recoil energies in the 100\,eV range; this technique should be also applicable to PALEOCCENE detectors.

\subsection{Threshold dependence on crystal direction}

Like all properties of a crystal, the threshold damage energy shows a dependence on the crystal direction. Simulation results for CaF$_2$ indicate a potential variation of this quantity by as much as a factor of 4 between the [100] and [111] directions~\cite{MORRIS2020109293}. Confirming this strong dependence on the crystal direction would allow to build detectors which can provide directional information: for CEvNS searches this would further help to suppress backgrounds and for dark matter searches it would allow to observe time variations in the signal. The period and phase of this time variation are well predicted by dark matter models and thus would help to establish the nature of the signal as being due to dark matter. To observe the dependence of threshold damage energy on the crystal direction, samples will be exposed to neutrons from the same source across a range of crystal directions and the response recorded.

\subsection{Gamma sensitivity}

It is well-known in the nuclear materials community that displacement damage can be produced not only by nuclear collisions but also by gamma-ray interactions. This is a potential background that needs to be carefully studied in experiments. Mosbacher {\it et al.}~\cite{Mosbacher:2019igk} experimentally studied the rate of color center formation for gammas and neutrons in several materials and identified several candidate materials for which this ratio is favorable for preferential nuclear recoil detection. Given the limited amount of information on this subject in the literature, a future research program should include systematic comparative studies of materials with gamma and neutron irradiation such that the impact of this background on the detectability of nuclear recoils can be better understood.

\subsection{Optimization of optical excitation}

The rate at which a crystal can be screened for the existence, location, and topology of color centers is dependent on efficacy of color center excitation that leads to optical emission. An ideal excitation source will be well aligned with the color center absorption spectrum and yield optical fluorescent emission with high probability. The screening study of Mosbacher {\it et al.}~\cite{Mosbacher:2019igk} used a monochromator-based source to evaluate several candidate crystals. Again, because of the limited availability of such studies in the literature, a dedicated study is needed that will determine the optical wavelength for the excitation source. One suitable method to do this is by use of a continuously tunable nanosecond to femtosecond optical parametric oscillator/amplifier.

\subsection{Simulating nuclear recoils with ions}

One difficulty with systematic studies of displacement damage cascades with neutrons is the stochastic nature of defect formation in the material volume and the ambiguities of the recoil angle, particularly at low energies. Other challenges include the practical availability of neutron sources in the required range of energies and the continuum of energy depositions that results from nuclear recoil, even when monoenergetic neutrons are used. To overcome these practical difficulties in experiment with neutrons, the nuclear materials community has used ion beams to simulate neutron damage and accelerate damage studies. This method had a significant potential in this context as well; specific ions species can be readily selected to emulate nuclear recoils, they can be focused to a specific area for easy post-analysis, their energy can be continuously tuned through a combination of accelerator voltage and thin attenuators placed in the beam, and they resemble a true monoenergetic spectrum of nuclear recoils. In addition, the available facilities offer a combination of irradiation with in-situ scanning electron microscopy that can reveal the dynamics of defect formation for subsequent comparison with the morphology of color centers. These fundamental studies are recommended as a component of a future R\&D program.

\section{Technical requirements}
\label{sec:tech}

The following discussion closely follows the appendix of Ref.~\cite{Cogswell:2021qlq} and full derivations can be found there. To scan a given volume per day, $V$, several factors need to be considered
\begin{enumerate}
    \item What is the density of pre-exisiting color centers?
    \item What laser power is need to achieve clear detection of color centers in the volume $V$?
    \item Is the resulting heat input into the sample manageable?
    \item Is the resulting data rate manageable?
\end{enumerate}

Light-sheet microscopes exist, for example~\cite{Tomer2015,mesoSPIM}, which can image 75$\,\mathrm{cm}^3$ per day. The results of Ref.~\cite{Mosbacher:2019igk} allow us to put an upper bound on the intrinsic color center density of  $f_d<10^{-12}-10^{-10}$. With these inputs we
obtain an estimate for the beam power of the exciting laser for a total volume of $1\,\mathrm{cm}^3$ and for $T=1$\,day:

\begin{eqnarray}
\label{eq:power}
    P&=&0.4\,\mathrm{W}\,\left(\frac{V_\mathrm{vox}}{\mu\mathrm{m}^3}\right)\left(\frac{10^6\,\mu\mathrm{m}^2}{\mathrm{fov}}\right)^{1/2}\left(\frac{f_d}{10^{-10}}\right)\left(\frac{\rho}{10\,\mathrm{g}\,\mathrm{cm}^{-3}}\right)
\left(\frac{1}{\xi}\right)\left(\frac{150\,\mathrm{g}\,\mathrm{mol}^{-1}}{M}\right)\times\nonumber\\ &&\quad\quad\quad\left(\frac{250\,\mathrm{nm}}{\lambda}\right) \left(\frac{1.5}{\mathrm{NA}}\right)^2\left(\frac{0.8}{\epsilon}\right) f\left(\frac{V_\mathrm{vox}}{V}\right)\,,
\end{eqnarray}

 where $P\propto V/T$ and $f(x)\simeq -0.0954 - 0.0912 \log_{10} x$, which accounts for counting statistics. In above equation $V$ is the volume of the crystal, $V_\mathrm{vox}$ is the volume of each voxel, $f_d$ is the pre-exisiting color center density in atoms per atom, $\xi$ is the oscillator strength of the color centers, $M$ is the molecular mass of the target, $\lambda$ is the excitation wavelength, $NA$ is the numerical aperture of the microscope objective in the imaging pathway and finally, $\epsilon$ is the quantum efficiency of the camera. In this derivation we assume that we detect the radiation induced color centers on top of the pre-existing ones by accumulating sufficient photon counting statistics per voxel to ensure that no more than 1 voxel in the entire volume $V$ on average is mis-reconstructed as having radiation induced color centers when in fact it does not. 
 
 Putting this amount of laser power into the crystal will result in a temperature rise of 
 \begin{equation}
 \label{eq:heat}
     \Delta T = 1\,\mathrm{K}\left(\frac{P}{\mathrm{W}}\right)\left(\frac{10\,\mathrm{W}\,\mathrm{K}^{-1}\,\mathrm{m}^{-1}}{\lambda}\right)\left(\frac{\alpha}{0.1\mathrm{cm}^{-1}}\right)\left(\frac{d}{\mathrm{cm}}\right)\,,
 \end{equation}
 where $\lambda$ is the heat conductivity of the crystal, $\alpha$ is the attenuation coefficient for the excitation wavelength and $d$ is the sample thickness. The amount of heating is small because our samples are essentially transparent and reasonable heat conductors.

The resulting data volume $B$ is is given as 
\begin{equation}
\label{eq:data}
    B=2\,\mathrm{TByte}\left(\frac{V}{1\,\mathrm{cm}^2}\right)\left(\frac{\mu\mathrm{m}^3}{V_\mathrm{vox}}\right)\left(\frac{b_d}{16\,\mathrm{bit}}\right)\,,
\end{equation}
 where $b_d$ is the bit depth of the image.
 
In summary, we find that neither the mechanics of scanning cubic centimeter volumes, the required laser power, the resulting heating of the crystal or data volume present major obstacles. One challenge could arise from the somewhat unusual wavelength requirements, in particular an illumination system in the deep UV, as indicated by the results of Ref.~\cite{Mosbacher:2019igk}, has not been demonstrated. However suitable components, in particular lasers and microscope objectives are available commercially.
\section{Summary}
\label{sec:summary}

The PALEOCCENE concept offers the potential for room-temperature, passive and robust detectors in the gram to kilogram range for the detection of low-energy nuclear recoil events. Nuclear recoil events can be caused by neutron scattering, CEvNS or dark matter scattering and therefore, PALEOCCENE could find applications in all three areas. In this contribution we outlined the current status and future plans for R\&D required to understand the feasibility and performance of detectors based on this concept.

\section*{Acknowledgements}
The Virginia Tech College of Science supported the work of
G. A. Khodaparast and R.H. H. Mudiyanselage with a  L. C. Hassinger Fellowship and the work of P. Huber with a R. Moore and  M. Khatam-Moore Fellowship. We are grateful to the Center for Microscopy and Image Analysis of the University of Zurich for the microscopy imaging, which was performed with their equipment and support. The work of P. Huber was supported by the U.S. Department
of Energy Office of Science under award number DE-SC00018327, the work of B. K. Cogswell and P. Huber by the 
National Nuclear Security Administration Office of Defense Nuclear
Nonproliferation R\&D through the Consortium for Monitoring,
Technology and Verification under award number DE-NA0003920, and the one of G. R. Araujo by the Candoc Grant No. K-72312-09-01 from the University of Zurich.

\bibliographystyle{apsrev-title}
\bibliography{apssamp.bib}

\end{document}